\documentclass[conference]{IEEEtran}
\IEEEoverridecommandlockouts
\usepackage{cite}
\usepackage{amsmath,amssymb,amsfonts}
\usepackage{algorithmic}
\usepackage{graphicx}
\usepackage{textcomp}
\usepackage{xcolor}
\usepackage[hang,small,bf]{caption}
\usepackage[subrefformat=parens]{subcaption}
 \usepackage{float}

\def\BibTeX{{\rm B\kern-.05em{\sc i\kern-.025em b}\kern-.08em
    T\kern-.1667em\lower.7ex\hbox{E}\kern-.125emX}}

\begin{document}
\title{Improvement of Serial Approach to Anomalous Sound Detection by Incorporating Two Binary Cross-Entropies for Outlier Exposure}

\author{\IEEEauthorblockN{Ibuki Kuroyanagi\IEEEauthorrefmark{1},
Tomoki Hayashi\IEEEauthorrefmark{1}\IEEEauthorrefmark{2},
Kazuya Takeda\IEEEauthorrefmark{1},
Tomoki Toda\IEEEauthorrefmark{1},
}
\IEEEauthorblockA{\IEEEauthorrefmark{1}Nagoya University, Nagoya, Japan}
\IEEEauthorblockA{\IEEEauthorrefmark{2}Human Dataware Lab. Co., Ltd., Nagoya, Japan}
\IEEEauthorrefmark{1}{\{kuroyanagi.ibuki, hayashi.tomoki\}@g.sp.m.is.nagoya-u.ac.jp},\\ \IEEEauthorrefmark{2}{hayashi@hdwlab.co.jp}, \IEEEauthorrefmark{1}{takeda@i.nagoya-u.ac.jp}, 
\IEEEauthorrefmark{1}{tomoki@icts.nagoya-u.ac.jp}
}

\maketitle

\begin{abstract}
Anomalous sound detection systems must detect unknown, atypical sounds using only normal audio data.
Conventional methods use the serial method, a combination of outlier exposure (OE), which classifies normal and pseudo-anomalous data and obtains embedding, and inlier modeling (IM), which models the probability distribution of the embedding.
Although the serial method shows high performance due to the powerful feature extraction of OE and the robustness of IM, OE still has a problem that doesn't work well when the normal and pseudo-anomalous data are too similar or too different.
To explicitly distinguish these data, the proposed method uses multi-task learning of two binary cross-entropies when training OE.
The first is a loss that classifies the sound of the target machine to which product it is emitted from, which deals with the case where the normal data and the pseudo-anomalous data are too similar.
The second is a loss that identifies whether the sound is emitted from the target machine or not, which deals with the case where the normal data and the pseudo-anomalous data are too different.
We perform our experiments with DCASE 2021 Task~2 dataset.
Our proposed single-model method outperforms the top-ranked method, which combines multiple models, by 2.1\,\% in AUC.
\end{abstract}

\begin{IEEEkeywords}
anomalous sound detection, outlier exposure, inlier modeling, hypersphere, multi-task learning
\end{IEEEkeywords}

\section{Introduction}
All machines in factories, plants, office buildings, etc., require regular maintenance to keep them functioning properly, and they can also break down or malfunction.
It is important to quickly address such equipment problems to prevent damage to the machine, or serious accidents.
In the past, skilled maintenance technicians would monitor the operation of machines and diagnose their condition by listening to them.
However, with the decrease in the rapidly aging working population, providing quality maintenance services with fewer skilled workers is becoming more challenging.
Furthermore, automated factories and plants are also becoming more common~\cite{huanh2018}. 
In response to these trends, methods for automatically detecting anomalous sounds have been developed~\cite{hayashi2018anomalous,bayram2021real}. 

Anomalous sound detection (ASD) is the task of identifying whether the sound emitted by a target machine is normal or anomalous. 
However, ASD is very different from simple, binary classification problems~\cite{Koizumi_DCASE2020_01} because it is difficult to collect data on every possible anomalous sound.
These sounds rarely occur during the normal operation, and the possible types of anomalous sounds are very diverse. 
Therefore, it is more practical to detect unknown, anomalous sounds using only normal sounds~\cite{kawaguchi2019}. 

Currently, two types of ASD approaches are mainly used: inlier modeling (IM) and outlier exposure (OE).
IM is a method that models the probability distribution of normal data and detects data that does not correspond to the model as anomalous data.
IM methods such as autoencoders~\cite{giri2020unsupervised,hayashi2020conformer,uefusa2020}, local outlier factor (LOF)~\cite{lof2000}, gaussian mixture models (GMM)~\cite{scott2004outlier,liu2019}, normalizing flows~\cite{Lopez2021,dohi2021} have been used.
IM is robust, but it is difficult to extract effective features.
In contrast, OE is a method for learning the decision boundaries of normal data by classifying normal and pseudo-anomalous data~\cite{primus2020anomalous,Miseul2021}.
OE methods such as deep semi-supervised anomaly detection~\cite{Ruff2020Deep,ruff2020rethinking}, and deep double centroids semi-supervised anomaly detection (DDCSAD)~\cite{kuroyanagi2021anomalous} have been used. 
OE is easy to extract effective features, but it is not robust. 
It also does not work well when normal and pseudo-anomalous data are too similar or too different~\cite{Kawaguchi2021}.

Recent studies have proposed methods that use a combination of IM and OE methods, and these approaches have achieved high ASD performance~\cite{Kawaguchi2021}.
One such approach, called the parallel method, ensembles anomaly scores for both IM and OE to compensate each other for the weakness of IM and OE~\cite{Lopez2021,Kuroyanagi2021dcasew}. 
However, the parallel method requires multiple models, which use different training processes, to be created to obtain combined anomaly scores, increasing the cost of system development and maintenance.
Another combination approach, called the serial method, uses OE- and IM-based methods in series.
The serial method consists of the following three steps:
First, we train the embedding with an OE method classification task using normal and pseudo-anomalous data.
Next, using the embedding extracted by the OE method in Step 1, an IM method trains a normal data distribution. 
Finally, data that differs from the distribution constructed in Step 2 is detected as anomalous.
The Serial method solves the difficulty of IM feature extraction and the robustness of OE~\cite{sohn2021learning,moritaSECOM2021,WilkinghoffFKIE2021}. 
Still, it leaves the problem of when the normal and pseudo-anomalous data of OE are too similar or too different.

To explicitly distinguish between normal and pseudo-anomalous data that are too similar or too different, the proposed method uses multi-task learning of two binary cross-entropies when training OE in the serial method.
Here, the datasets~\cite{mimii2021, Harada2021} contain several machine types such as fan, pump, valve, etc. 
Each machine type has $K$ product IDs such as ID 0, 1, 2, etc.
The first is a loss that classifies the sound of the target machine type to which product ID it is emitted from, which deals with the case where the normal data and the pseudo-anomalous data are too similar.
The second is a loss that identifies whether the sound is emitted from the target machine type or not, which deals with the case where the normal data and the pseudo-anomalous data are too different.
The proposed method overcomes the weaknesses of OE. 
Furthermore, the proposed method of OE is easy to develop and manage since one model corresponds to one machine type, and there is no need to create a model for each product ID~\cite{Kuroyanagi2021dcasec}.
We evaluate the performance of the proposed method by conducting experiments and showing its effectiveness. 
Also, we visualize the embedding by t-SNE~\cite{t_sne} and qualitatively discuss the effect of using the serial method and the change of embedding by the proposed method.

\begin{figure}[htbp]
  \begin{minipage}[b]{0.47\linewidth}
    \centering
    \includegraphics[keepaspectratio, scale=0.11]{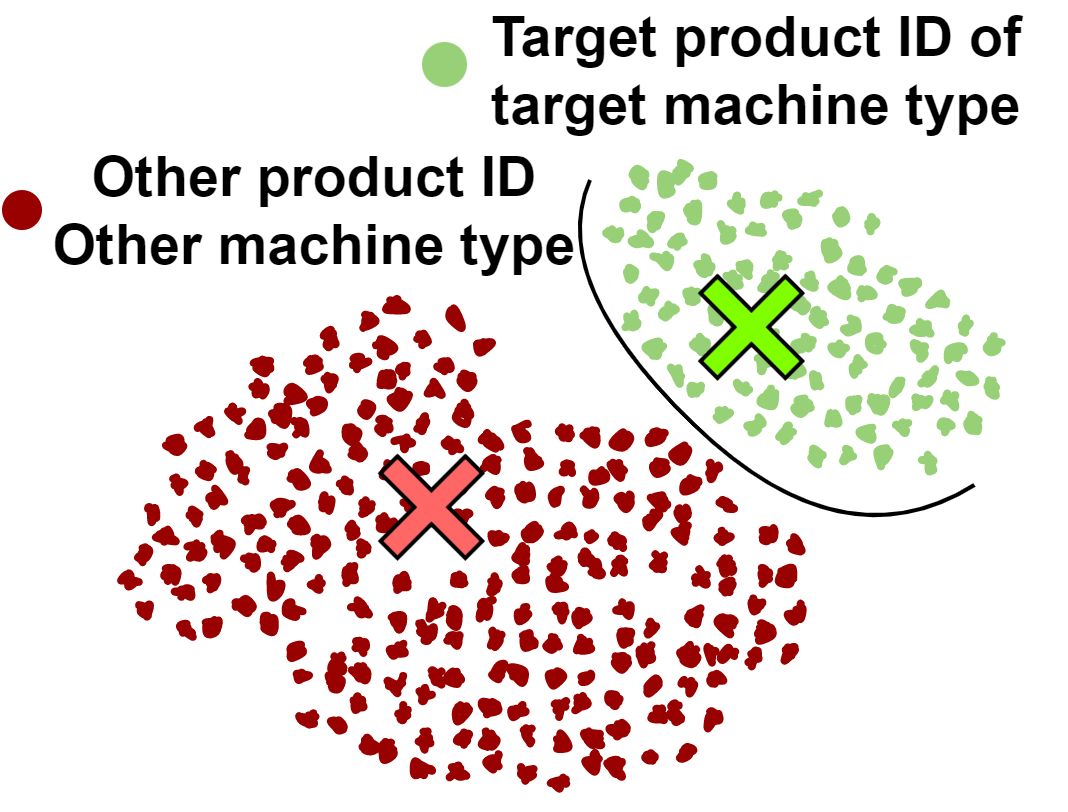}
    \subcaption{DDCSAD}
    \label{fig:embed_ddcsad}
  \end{minipage}
  \begin{minipage}[b]{0.47\linewidth}
    \centering
    \includegraphics[keepaspectratio, scale=0.125]{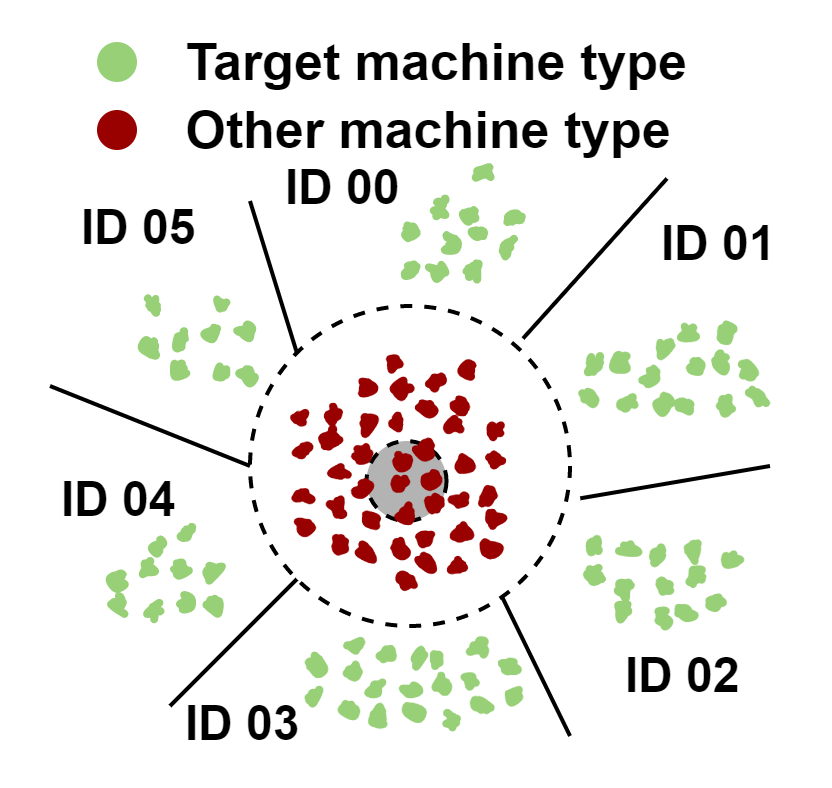}
    \subcaption{Proposed method}
    \label{fig:embed_proposed}
  \end{minipage}
  \caption{Images of embeddings which obtained by each method.}
\end{figure}

\section{OE-based method}
\label{related}
To improve the OE component of our serial method, we focus on DDCSAD~\cite{kuroyanagi2021anomalous}, which has demonstrated high ASD performance.
DDCSAD is trained using multi-task binary classification and metric learning. 
A binary classification model is trained using the target product ID of the target machine type as positive samples, and the other product IDs and the other machine types as negative samples.
In addition, we define the centroid for positive and negative samples and perform metric learning to minimize within-class variance and maximize between-class variance for each class.
During inference, we use the probability calculated using binary classification and the weighted average of the distance between the centroid of the positive samples and the embedding as the anomaly score. 

Fig.~\ref{fig:embed_ddcsad} shows an image of an embedding obtained using DDCSAD.
In this method, the pseudo-anomalous data move away from the centroid of the positive sample and move closer to the negative sample, which is considered robust when the distributions of normal and anomalous data are very different.
However, there are several problems with this method.
First, when the distributions of normal and pseudo-anomalous data are similar, the performance of this method is degraded because the decision boundaries of normal and pseudo-anomalous data are different from those of normal and anomalous data.
Second, when using Euclidean distance, the embedding of normal data is assumed to follow a normal distribution. However, if the model is not expressive enough, the above assumption cannot be satisfied, and performance is degraded.
Furthermore, since a model is created for each product ID, the performance variation of the model becomes large. In addition, the development and maintenance of the model become complicated.

\section{Proposed Method}
\begin{figure}[thb]
  \centering
  \centerline{\includegraphics[width=1\columnwidth]{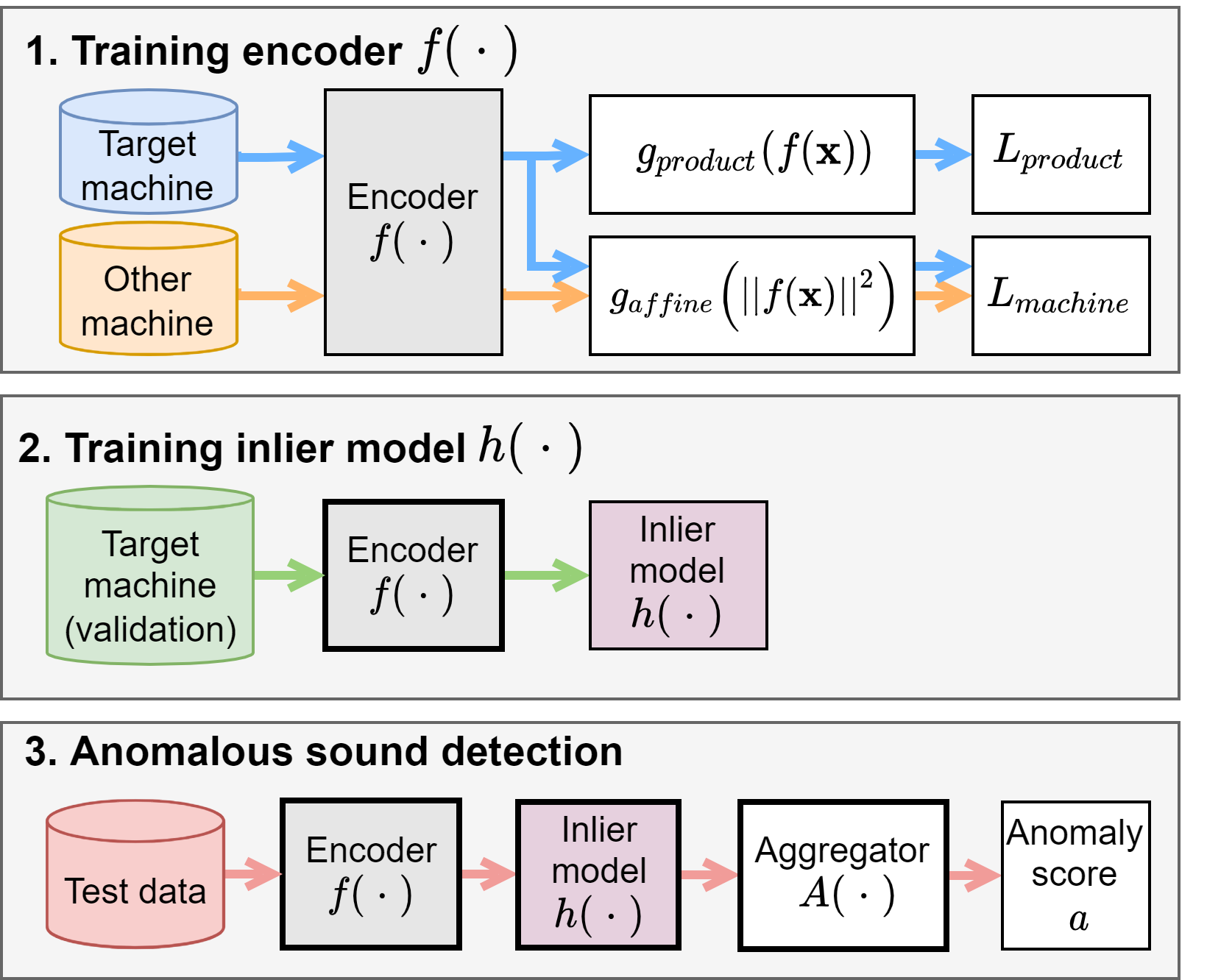}}
  \caption{Overview of our proposed method.}
  \label{fig:proposed_method}
\end{figure}

Fig.~\ref{fig:embed_proposed} shows an image of an embedding obtained by our proposed method.
Fig.~\ref{fig:proposed_method} shows an overview of the proposed method. 
To explicitly distinguish between normal and pseudo-anomalous data that are too similar or too different, the proposed method uses multi-task learning of two binary cross-entropies when training OE.
The first is a product ID classification loss, which deals with cases where the normal and pseudo-anomalous data are too similar.
The second is a machine type classification loss, which deals with cases where the normal and pseudo-anomalous data are too different.

If the audio input is represented as $x_i~(i=1,2,...,N)$, the machine type is represented as $t_i~(i=1,2,...,N)$, where $t_i$ is 1 for the target machine type and 0 for the other machine types.
Each machine type has $K$ product IDs, and $x_i$ belongs to one of them.
The one-hot vector for the product ID is represented as
$y_{\{i,k\}}~(i=1,2,...,N,k=1,2,...,K)$, where $y_{\{i,k\}}$ is 1 for the $k$~th element and 0 for the other elements when the product ID is $k$.
Product IDs classification loss can be represented as:
\begin{equation}
    \scalebox{0.96}{$\displaystyle
    \label{eq:l_product}
    \begin{split}
        \mathcal{L}_{\rm product} &= -\frac{1}{K\sum_{i=1}^Nt_i }\sum_{i=1}^N\sum_{k=1}^Kt_i\left\{y_{\{i,k\}}{\rm log}\left(g_{\rm product}\left(f(x_i)\right)\right)\right.\\
        &\left.+(1-y_{\{i,k\}}){\rm log}\left(1-g_{\rm product}\left(f(x_i)\right)\right)\right\},
    \end{split}
    $}
\end{equation}
where $f$ is an encoder, and $g_{\rm product}$  is a linear transformation. 
Machine type classification loss can be represented as:
\begin{equation}
    \label{eq:l_machine}
    \begin{split}
        \mathcal{L}_{\rm machine} &= -\frac{1}{N}\sum_{i=1}^N\left\{ t_i{\rm log}\left(g_{\rm affine}\left(||f\left(x_i\right)||^2\right)\right) \right.\\
        &\left.+(1-t_i){\rm log}\left(1-g_{\rm affine}\left(||f\left(x_i\right)||^2\right)\right)\right\},
    \end{split}
\end{equation}
where $g_{\rm affine}$ is an affine transformation. 
When creating mini-batches, we use a batch sampler so that the value of $t$ is 1:1. 
The final combined loss function is:
\begin{equation}
    \label{eq:l_final}
    \begin{split}
        \mathcal{L} = \mathcal{L}_{\rm machine}+\lambda \mathcal{L}_{\rm product},
    \end{split}
\end{equation}
where $\lambda$ is a hyperparameter.
Mixup~\cite{mixupzhang2018} is applied in mini-batches to obtain intermediate features between normal and pseudo-anomalous data, and each sample applies Eq.~\ref{eq:l_product} if the target machine type is included. 
In other words, we identify the product IDs of the target machine type but not the product IDs of the other machine types.
The method for calculating the anomaly score $a_i~(i=1,2,...,N)$ is:
\begin{equation}
    \label{eq:anomaly_score}
    \begin{split}
        a_i = \mathcal{A}\left(h\left(f\left(\mathcal{X}_i\right)\right)\right),
    \end{split}
\end{equation}
where $\mathcal{X}_i$ is a set of $S$ segments that divide $x_i$ into $T$ seconds, allowing for overlap, $h$ is a post-processing function for IM such as GMM or LOF, and $\mathcal{A}$ is the aggregator of the anomaly scores such as max\,/\,average pooling.

By using the product IDs classification, anomalous sounds that are similar to normal sounds are distributed around each product ID.
By using the machine type classification by the norm of embedding, anomalous sounds that differ significantly from normal sounds are collected near the center of the hypersphere.
This approach is inspired by DDCSAD’s training method, which involves collecting pseudo-anomalous data into a single point.
Since the proposed method explicitly distinguishes between normal and pseudo-anomalous data that are too similar or too different, it avoids the insufficient expressiveness of the OE embedding.
In other words, the proposed method obtains an embedding suitable for detecting normal or similar sounds.
In addition, it is easy to develop and maintain the system because the model is created for each machine type, not for each product ID in each machine type.  

\begin{table}[htb]
\centering
\caption{{Hyperparameters for each machine type.}\\{LR=learning rate, BS=batch size}}
\label{table:hyperparams}
\scalebox{0.89}{$\displaystyle
\begin{tabular}{c|ccccccc}
\hline
\multicolumn{1}{l|}{} & fan & gearbox & pump & valve & slider & ToyCar & ToyTrain \\ \hline
LR & $0.001$ & $0.001$ & $0.001$ & $0.0005$ & $0.001$ & $0.001$ & $0.0005$ \\
BS & 32 & 128 & 128 & 128 & 128 & 128 & 32 \\
$\lambda$ & 0.1 & 10.0 & 10.0 & 10.0 & 10.0 & 10.0 & 0.1 \\
$h$ & GMM & GMM & GMM & GMM & GMM & LOF & LOF \\
$p$ & 16 & 64 & 2 & 32 & 2 & 16 & 8 \\ \hline
\end{tabular}
$}
\end{table}

\begin{figure}[htbp]
    \begin{tabular}{cc}
      \begin{minipage}[t]{0.47\hsize}
        \centering
        \includegraphics[keepaspectratio, scale=0.31]{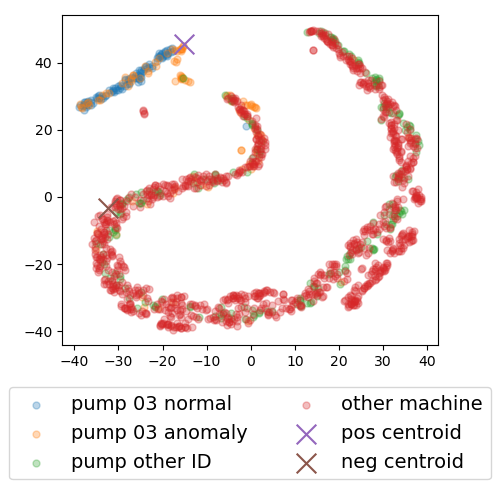}
        \subcaption{DDCSAD pump ID 03}
        \label{ddcsad_pump_p29}
      \end{minipage} &
      \hspace{-0.1\columnwidth}
      \begin{minipage}[t]{0.47\hsize}
        \centering
        \includegraphics[keepaspectratio, scale=0.31]{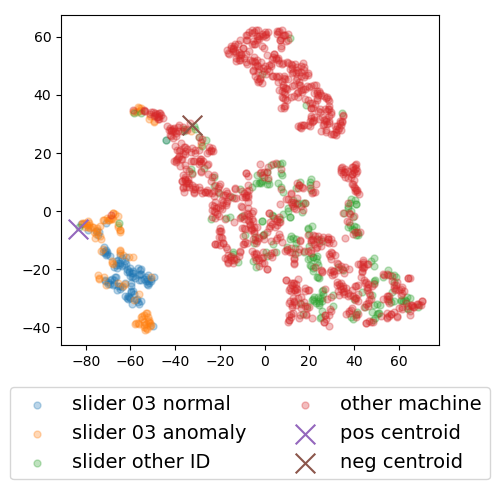}
        \subcaption{DDCSAD slider ID 03}
        \label{ddcsad_slider_p17}
      \end{minipage} \\ 
      
      \begin{minipage}[t]{0.47\hsize}
        \centering
        \includegraphics[keepaspectratio, scale=0.31]{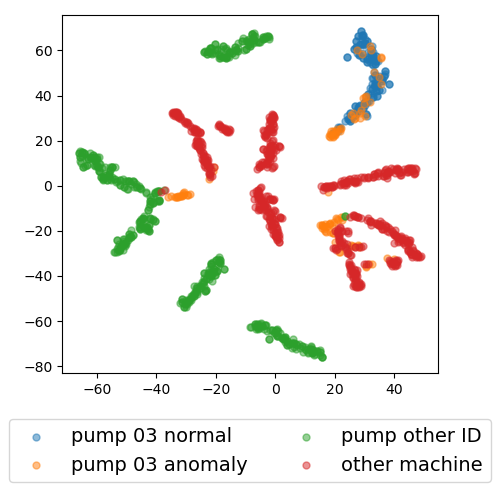}
        \subcaption{Proposed method \\pump ID 03}
        \label{proposed_pump_section03_p18}
      \end{minipage} &
      \hspace{-0.1\columnwidth}
      \begin{minipage}[t]{0.47\hsize}
        \centering
        \includegraphics[keepaspectratio, scale=0.31]{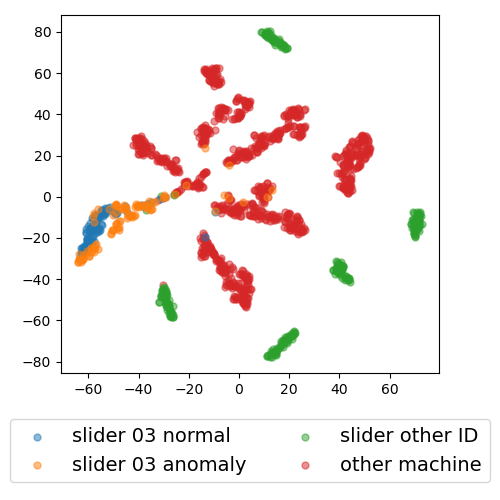}
        \subcaption{Proposed method \\slider ID 03}
        \label{proposed_slider_section03_p17}
      \end{minipage} \\
      
      \begin{minipage}[t]{0.47\hsize}
        \centering
        \includegraphics[keepaspectratio, scale=0.31]{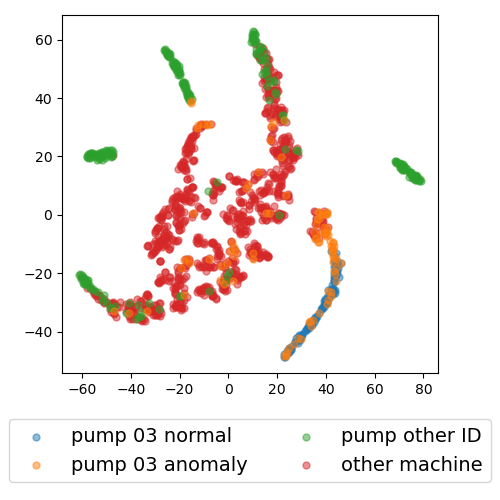}
        \subcaption{Only product IDs \\classification pump ID 03}
        \label{wo_pump_section03_p20}
      \end{minipage} &
      \hspace{-0.1\columnwidth}
      \begin{minipage}[t]{0.47\hsize}
        \centering
        \includegraphics[keepaspectratio, scale=0.31]{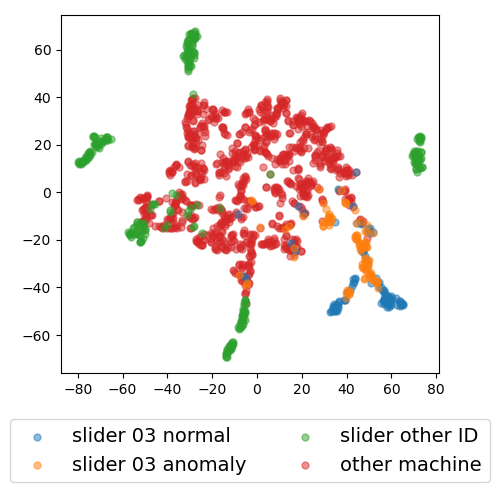}
        \subcaption{Only product IDs \\classification slider ID 03}
        \label{wo_slider_section03_p15}
      \end{minipage}
    \end{tabular}
    \caption{Visualizations of various embeddings}
    \label{fig:embed}
\end{figure}

\section{Experimental evaluation}
\begin{table*}[htbp]
\centering
\caption{Performance evaluation results. Values represent the harmonic mean of AUC [\%] and pAUC ($p = 0.1$) [\%] for each product ID.\\“All\,/\,Har-mean” column values represent the harmonic mean of AUC and pAUC over all machines and product IDs.\\ $\#$models column values represent the number of required models for one machine type. 
}
\label{table:result1}
\begin{tabular}{c|ccccccc|c|c}
\hline
Method & fan & gearbox & pump & valve & slider & ToyCar & ToyTrain & All$\,$/$\,$Har-mean &$\#$models\\ \hline
Parallel method~\cite{Lopez2021} & 56.11 & 62.43 & 85.33 & 66.68 & 74.58 & 68.46 & 69.04 & 67.93& 5\\\hline
Serial method~\cite{moritaSECOM2021} & 82.37 & 66.22 & \textbf{77.21} & 72.06 & 78.02 & 54.72 & 50.88 & 66.78 & 1\\
DDCSAD~\cite{kuroyanagi2021anomalous} & 69.21 & 58.13 & 68.55 & \textbf{75.66} & 59.56 & 57.71 & 57.71 & 63.12 & 3\\
Proposed method& \textbf{84.35} & \textbf{68.42} & 71.60 & 65.03 & \textbf{83.97} & \textbf{62.08} & \textbf{58.83} & \textbf{69.42} & 1\\ \hline
\end{tabular}
\end{table*}

\begin{table*}[htbp]
\centering
\caption{Ablation study results. Values represent the harmonic mean of AUC [\%] and pAUC ($p = 0.1$) [\%] for each product ID.\\
 “All\,/\,Har-mean” column values represent the harmonic mean of AUC and pAUC over all machines and product IDs.}
\label{table:result2}
\begin{tabular}{c|ccccccc|c}
\hline
Method & fan & gearbox & pump & valve & slider & ToyCar & ToyTrain & All$\,$/$\,$Har-mean \\ \hline
Proposed method& 84.35 & 68.42 & \textbf{71.60} & 65.03 & \textbf{83.97} & \textbf{62.08} & 58.83 & \textbf{69.42} \\
w/o Mixup & \textbf{86.35} & \textbf{71.54} & 66.52 & \textbf{67.66} & 81.88 & 52.51 & \textbf{59.90} & 67.75 \\
w/o $h(\cdot)$ & 64.34 & 49.80 & 66.25 & 57.19 & 61.69 & 50.51 & 53.35 & 56.93 \\
Only product IDs classification & 85.39 & 66.07 & 66.88 & 58.16 & 70.50 & 51.68 & 52.78 & 62.79 \\ 
\hline
\end{tabular}
\end{table*}
\subsection{Experimental conditions}
To evaluate the performance of the proposed method, we conducted experiments using the data from the DCASE 2021 Task~2 Challenge~(MIMII Due\cite{mimii2021}, ToyADMOS2~\cite{Harada2021}). 
The training and evaluation data in the same domain were used.
We used sound from seven machine types: fan, gearbox, pump, valve, slider, ToyCar, and ToyTrain. 
Each machine type has 6 product IDs.
For each product ID, 1,000 normal sound samples were used as training data, while 100 normal and 100 anomalous sound samples were used for evaluation data.
ID 0, 1, and 2 of the evaluation data were used as validation data to determine the hyperparameters $p$ of IM and $\lambda$, and ID 3, 4, and 5 were used as test data. 
When training Encoder $f$, 90\,\% of the training data was randomly selected, and the remaining 10\,\% was used for validation.
Each recording is a single-channel, 10 sec. segment of audio sampled at 16\,kHz.

Each machine’s amplitude was calculated and normalized during preprocessing to have a mean of 0 and a variance of 1.
For each audio input sequence, we extracted a log-compressed Mel-spectrogram with a window size of 128\,ms, a hop size of 16\,ms, and 224 Mel-spaced frequency bins in the range of 50--7800\,Hz, in 2.0\,sec.
These features were passed to encoder $f$ using EfficientNet-B0~\cite{xie2020self}.
Encoder $f$ applied global average pooling to the last convolutional layer and performs two non-linear transformations to obtain a 128-dimensional embedding.
Weights of $g_{\rm product}$ and $g_{\rm affine}$ were handled as trainable parameters.
We used AdamW~\cite{adamw2019}, OneCycleLR~\cite{onecyclelr2019}, and Mixup ($\alpha=0.2$) for training. 
Learning rate, batch size, $\lambda$, inlier model $h$, and the inlier model's hyperparameters $p$, which the number of components for GMM and the number of neighbors for LOF are shown in Table~\ref{table:hyperparams}.
We used the model with the smallest loss of validation data of the training data after 300 epochs. 
We trained the inlier model $h$ using the validation data of the training data for each product ID.
We determined the hyperparameters of the inlier model $h$ using the validation data of the evaluation data.
During inference, we divided 10.0\,sec. clips into $S=10$ segments, with overlap allowed so that each segment was $T=2.0$\,sec.
The GMM used the negative log-likelihood as the anomaly score, while the LOF used the outlier score. 
The aggregator $\mathcal{A}$ was the mean of the anomaly scores above the median for the GMM and the mean of the entire anomaly scores for the LOF. 

\subsection{Results}
The parallel~\cite{Lopez2021} and serial~\cite{moritaSECOM2021} methods shown in Table~\ref{table:result1} are the first and second ranked methods in the DCASE 2021 Task~2 Challenge, respectively, while DDCSAD~\cite{kuroyanagi2021anomalous} is a conventional method.
Table~\ref{table:result1} shows that for All$\,$/$\,$Har-mean, the proposed method outperformed all of the other methods.
As shown in the $\#$models in Table~\ref{table:result1}, the proposed method using a single model outperformed the parallel method ensembled different five models.
The proposed method improved the performance of ToyCar and ToyTrain by more than 7\,\% over the serial method, and we believed that the proposed method obtained more suitable embedding for ASD.
We believe that the reason for the performance improvement of the proposed method compared to DDCSAD is that it is more flexible and effective in modeling the normal data by IM without assuming a normal distribution for the normal data.
Focusing on $h$, we achieved the best performance when using a GMM with the machine types in MIMII Due dataset and a LOF with the machine types in ToyADMOS2 dataset, suggesting that it was important to use a suitable function $h$ for the dataset. 

Our ablation study results are shown in Table~\ref{table:result2}. 
Fig.~\ref{fig:embed} shows the results when the embedding of each method is visualized using t-SNE~\cite{t_sne}.
Table~\ref{table:result2} shows that using Mixup improved the pump, slider, and ToyCar.
We believed that Mixup effectively trained data with a high similarity between product IDs.
However, Mixup did not improve ASD performance for the other machine types.
Based on these, we considered Mixup to be one of the most important hyperparameters.

We then examined the difference in performance with and without $h$, the IM component of our proposed method. 
When we did not use $h$, we used the average of the output probabilities of the product IDs as the anomaly score instead~\cite{Kawaguchi2021}. 
The experiment shows that IM was effective for ASD. 
We believe that even though some parts of the data distribution are less informative for the product ID classification and tend to be ignored in calculating the output probabilities of the product IDs, they are still helpful in detecting differences in the data distribution between normal and anomalous data. 
IM can detect such differences in distribution by GMM or LOF.

We then considered the effect of Eq.~\ref{eq:l_machine}, which calculates the machine type classification loss. 
The only product IDs classification in Table~\ref{table:result2} is almost the same as the serial method~\cite{moritaSECOM2021}.
The performance of the proposed method outperformed that of the only product IDs classification, since Eq.~\ref{eq:l_machine} collected anomalous data that differed significantly from normal data near the center of the hypersphere.
More anomalous data was distributed near the other machine types when using the proposed method, as shown in Fig.~\ref{proposed_pump_section03_p18} and Fig.~\ref{proposed_slider_section03_p17}, than when the only product IDs classification was used, as shown in Fig.~\ref{wo_pump_section03_p20} and Fig.~\ref{wo_slider_section03_p15}.

\section{Conclusion}
In this paper, we proposed using multi-task learning of two binary cross-entropies: the product ID classification loss and the machine type classification loss when training OE in the serial method.
The product ID classification loss and the machine type classification loss correspond when normal and pseudo-anomalous data are too similar or too different, respectively.
Even though the proposed method used a single model, it outperformed conventional ensembled methods.
Our ablation study and visualization of embeddings confirmed the effectiveness of using post-processing and the effectiveness of the proposed method when detecting anomalous data that is too different from normal data.
In future work, we will investigate the impact of training data on ASD performance and the use of data with different domains.
\section*{Acknowledgment}
This paper was partly supported by a project, JPNP20006, commissioned by NEDO.

\bibliographystyle{IEEEtran}
\bibliography{ref}
\end{document}